\documentclass[aps,prd,showpacs,amsmath,amssymb]{revtex4}
\usepackage[]{latexsym}
\usepackage[english]{babel}
\usepackage{graphicx}
\usepackage{hyperref}
\usepackage{color}
\renewcommand{\d}{\mathrm{d}}

\def\be{\begin{equation}}
\def\ee{\end{equation}}
\def\bea{\begin{eqnarray}}
\def\eea{\end{eqnarray}}

\begin{document}

\title{Dark energy from Geometrothermodynamics}

\author{Alessandro Bravetti}
\email{bravetti@icranet.org}
\affiliation{Dipartimento  di Fisica and Icra, "Sapienza" Universit\`a di Roma,
            Piazzale Aldo Moro 5, I-00185, Roma, Italy}
\affiliation{Instituto de Ciencias Nucleares, Universidad Nacional
Aut\'onoma de M\'exico, AP 70543, M\'exico, DF 04510, Mexico.}

\author{Orlando Luongo}
\email{orlando.luongo@na.infn.it}
\affiliation{Dipartimento di Fisica, Universit\`a di Napoli "Federico II",
             Via Cinthia, I-80126, Napoli, Italy;}
\affiliation{INFN Sez. di Napoli, Monte Sant'Angelo, Edificio N,
Via Cinthia, I-80126, Napoli, Italy}
\affiliation{Instituto de Ciencias Nucleares, Universidad Nacional
Aut\'onoma de M\'exico, AP 70543, M\'exico, DF 04510, Mexico.}

\date{\today}

\begin{abstract}
Geometrothermodynamics is a geometric theory which combines thermodynamics with contact and Riemannian geometry.
In this work we use the formalism of geometrothermodynamics to infer cosmological models which predict the observed speed up. As a relevant consequence, our simple model shows dynamical properties which seem to fairly well describe the late time universe dynamics. To do so, we use geometric considerations about constant thermodynamic curvature and derive the model of a fluid which is expected to \emph{naturally} reproduce the dark energy effects. In particular, our  approach reduces to the $\Lambda$CDM model in the limiting case of small redshift, providing however significative departures from $\Lambda$CDM as the universe expands. The main goal consists in interpreting our \emph{geometrothermodynamic fluid} as an energetic source and to explain the dark energy effects as emerging from
 the interplay between geometry and thermodynamics, providing a new interpretation of the observed positive acceleration.
\end{abstract}

\pacs{04.50.-h,  04.20.Cv, 98.80.Jk}

\maketitle

\section{Introduction}

Geometrothermodynamics (GTD) is a recent theory that combines differential geometry and thermodynamics \cite{quev07}.
The main aim of GTD is to describe the equilibrium states of a thermodynamic system as points of the so-called \emph{equilibrium space}, hereafter ${\cal E}$. To define such a space, GTD requires the definition of a Riemannian metric $g$, which plays the role of thermodynamical ruler between states. According to the physical properties of thermodynamics, such a metric structure is chosen to be invariant under Legendre transformations of the potential  \cite{quev07}. It results that  the metric structure $g$ and  the geometric properties of ${\cal E}$ are fully determined once the fundamental equation for the thermodynamic potential is explicitly  known, as it happens in standard thermodynamics. It is then expected that the thermodynamic properties of the system can be represented in terms of the geometric properties of ${\cal E}$. In particular, the scalar curvature of  ${\cal E}$ takes account of the thermodynamic interaction and therefore a flat thermodynamic geometry corresponds to ideal gases, while curvature singularities to phase transitions \cite{hernando2, PhTransGTD, AppNewDev}. Once the geometric paradigm for doing thermodynamics is established, it turns out that one can use geometric tools to work out new useful thermodynamic relations. For example, one can generate fundamental equations that determine interesting geometric properties, such as e.g. extremal thermodynamical surfaces or spaces of constant thermodynamic curvature. In the context of cosmology, the former approach has been shown to extend well known fluids, such as the Chaplygin gas and the modified Chaplygin gas (see \cite{AlejQuev}), while the latter technique has been recently investigated in \cite{AppNewDev,EinsteinM} and it also provides fundamental equations that can coherently reproduce the behavior of the dark sector of the universe \cite{oncia}.

In this work we derive a new cosmological model by means of GTD. In particular, we combine heuristic thermodynamic and geometric considerations in the GTD paradigm and find out a novel equation of state which results as geometrical problem solution and that correctly describes the universe expansion today in the context of Friedmann-Robertson-Walker cosmology. In fact, our model can explain recent cosmological observations \cite{SNeIa1,SNeIa2,SN21,SN22,SN23}, which definitively showed that the universe is currently undergoing an accelerated expansion. In particular, to characterize the net energy budget of the universe, a dark energy (DE) component is derived from GTD, through the introduction of a \emph{geometrothermodynamic fluid} (GTDF), which is supposed to drive the positive acceleration. The main property of the GTDF is that it exhibits a negative equation of state, able to counterbalance the positive attraction due to gravity \cite{boliok,cop}. In addition, such a behavior seems to naturally emerge only at late times during the universe expansion, showing that the DE contribution may be neglected after a certain redshift, namely the \emph{transition redshift}, while it becomes significative at our time \cite{transix}. The basic argument of our approach is that the GTDF exhibits a pressure proportional to the volume occupied by the fluid itself, giving rise to a simple explanation of the emergence of DE effects: during the universe evolution the volume increases and therefore the role of DE becomes more important, since its (negative) pressure increases as the volume. This turns out to be a consequence of assuming a constant thermodynamic interaction.
In the context of GTD we can argue a precise definition for the concept of "thermodynamic interaction'',
by postulating that the interaction is represented by curvature of the equilibrium manifold. Thus, our assumption is rephrased by requiring the equilibrium manifold to have constant scalar curvature. Given the aforementioned assumptions, we construct our model, deriving the DE fundamental equation in terms of the GTDF. Interestingly, we will show that our GTDF is characterized by an evolving equation of state, whose limiting case reduces to a cosmological constant, with constant and negative pressure. This leads to a self-consistent description of the late time universe dynamics, framing the role played by DE as due to thermodynamical interactions.

This paper is organized as follows. In Sec. \ref{basicGTD} we review the basics of GTD, focusing on the introduction of the Riemannian structure on the equilibrium manifold. In Sec. \ref{cosmoGTD} we find a particular solution to the problem of constant thermodynamic curvature in the GTD context and show that the resulting fundamental thermodynamic equation can have interesting properties in the context of cosmology. Later in Sec. \ref{FRWmodel} such properties are investigated in detail assuming a FRW cosmology and in Sec. \ref{LCDM} we show that our model reduces to $\Lambda$CDM for small redshifts. Finally, in Sec. \ref{Conclusion} we outline the conclusions and suggest further perspectives.

\section{Geometrothermodynamics}\label{basicGTD}

In this section, we provide the basic aspects of Geometrothermodynamics (GTD). 
The formalism of GTD is based on the core idea that the equilibrium points of a thermodynamic system can be described in terms of a Riemannian manifold,
i.e. the equilibrium manifold $(\mathcal E, g)$.  Within this scheme, the thermodynamic interaction is represented by  the scalar curvature of $(\mathcal E, g)$.
Since the aim is to describe thermodynamic systems, GTD promotes to choose the thermodynamic metric $g$
in such a way that it results to be invariant under Legendre transformations (LTs), which play a central role in ordinary homogeneous thermodynamics \cite{Callen}.
To do so, in GTD one needs to first introduce an ambient manifold, called the \emph{phase manifold},  which is useful to correctly define and handle the LTs, in a way we will now specify.

\subsection{The phase manifold $(\mathcal T,\Theta,G^\natural)$}

Given a thermodynamic system with $n$ degrees of freedom, the {\it thermodynamic phase manifold} is the $(2n+1)$-dimensional  manifold $\mathcal{T}$,
endowed with a contact structure $\xi\subset T\mathcal{T}$, that is, a maximally non-integrable family of hyperplanes of the form $\xi=\ker(\Theta)$,
where $\Theta$ is a $1$-form, fulfilling the non-integrability condition

\begin{equation}\label{integraxx}
   \Theta \wedge (\d \Theta)^n \neq 0\,.
\end{equation}

\noindent It turns out (see  e.g. \cite{Arnold}) that one can always find a set of local coordinates on $\mathcal{T}$, which we indicate as
\begin{equation}\label{aluusjgg}
Z^A =
(\Phi, E^1, \dots, E^n, I_1, \dots, I_n)\,,
\end{equation}
such that the contact $1-$form $\Theta$ may be rewritten as
\begin{equation}\label{ardnol}
\Theta = \d \Phi - I_a \d E^a\,.
\end{equation}
In the phase space $\mathcal T$, the Legendre transformations (LTs)
are special changes of coordinates given by
\begin{equation}
\{\Phi, E^a, I^a\} \longrightarrow \{\tilde \Phi, \tilde E ^a, \tilde I ^ a\}\ ,
\end{equation}
\begin{equation}
 \Phi = \tilde \Phi - \delta_{kl} \tilde E ^k \tilde I ^l \ ,\quad
 E^i = - \tilde I ^ {i}, \ \
E^j = \tilde E ^j,\quad
 I^{i} = \tilde E ^ i , \ \
 I^j = \tilde I ^j \ ,
 \label{leg}
\end{equation}
where  $i\,\cup j$ represent any disjoint decomposition of the set of indices $\{1,...,n\}$, and $k,l= 1,...,i$.
The peculiarity of LTs is that they leave the contact structure $\xi$ invariant. As such they belong to the family of {\it contactomorphisms} (see e.g. \cite{Arnold}). Another appealing class of transformations in thermodynamics are the ones which interchange the representation, for instance from the entropy to the internal energy or viceversa. It can be shown that such transformations are also contactomorphisms and they can be written as changes of coordinate in $\mathcal T$ (c.f.  \cite{NewDev} for the full discussion) as
	\begin{equation}
	\label{CR2}
	\Phi' = E^{(i)}, \quad
	E^{(i)'}  =  \Phi, \quad
	E^{j'} =  E^j, \quad
	I_{(i)'}  =  \frac{1}{I_{(i)}} \quad \text{and} \quad
	I_{j'} =  -\frac{I_j}{I_{(i)}}\,,
	\end{equation}
where we use an index $(i)$ to specify the variable which is chosen to be the new thermodynamic potential in the change of representation (usually $\Phi=U$ and
$E^{(i)}=S$ or viceversa). It was shown in \cite{NewDev} that it is possible to find a metric structure in $\mathcal T$ that has the LTs and the change of representation from entropy to energy as isometries.
It was proposed that such a metric should be the natural metric to study thermodynamics in the GTD context. In the local coordinates defined in (\ref{aluusjgg}), such a metric reads
\begin{equation}  \label{Gnat}
G^{\natural} = \Theta \otimes \Theta +\sum_{j \neq i} \frac{1}{E^j I_j} \d E^a \otimes \d I_a,
\end{equation}
where in the sum we excluded the $i$th pair of coordinates which is to be exchanged for the thermodynamic potential $\Phi$ when changing from one representation to another,  as in (\ref{CR2}).

\subsection{The equilibrium manifold $(\mathcal E,g^\natural)$}

Having introduced the thermodynamic phase space,
in GTD one defines the {\it equilibrium space} as the maximum integral sub-manifold of $\mathcal T$,
i.e. the embedded manifold given by the map
\begin{equation}  \label{embedding}
\varphi: \mathcal{E} \to \mathcal{T},
\end{equation}
where the isotropic condition $\varphi^*(\Theta) = 0$ holds. It turns out that $\mathcal E$ is an $n$-dimensional
manifold defined as the set for which the coordinate functions of $\mathcal{T}$ satisfy the conditions
	\begin{equation}    \label{first}
	\d \Phi - I_a \d E^a = 0, \qquad {\rm and} \qquad I_a = \frac{\partial \Phi}{\partial E^a}.
	\end{equation}
Equations \eqref{first}  are the first law of thermodynamics and the set of equations of state, respectively.
It is then clear that $\Phi(E^a)$ is the fundamental relation for the thermodynamic system,
$E^a$ are the extensive variables and $I_a$ their corresponding intensive parameters.
It is then easy to give $\mathcal E$ a Riemannian structure. This is performed by using the pullback of the metric structure $G$ of $\mathcal T$, via the map $\varphi$.
Given the Riemannian structure for the phase space in (\ref{Gnat}), it follows that the induced Riemannian structure in $\mathcal E$ is given by
\begin{equation}   \label{gnat}
g^\natural  = \sum_{j \neq i} \left( E^j \frac{\partial \Phi}{\partial E^j} \right)^{-1} \frac{\partial^2 \Phi}{\partial E^b \partial E^a}\, \d E^a \otimes \d E^b.
\end{equation}

\noindent We remark here that the geometric counterpart of the thermodynamic systems is the equilibrium manifold ${\cal E}$ and the scalar
curvature of ${\cal E}$ is conjectured to be an invariant measure of the thermodynamic interaction.
It follows, for example, the intriguing fact that curvature singularities of ${\cal E}$ represent phase transitions \cite{PhTransGTD, AppNewDev, HighBH, Unified}
and the geodesics of ${\cal E}$ represent quasi-static processes \cite{quasistatic, Diego}. Furthermore, it was also pointed out that equations of state describing the dark sector
of the universe dynamics can be translated in this geometric language and they exhibit some interesting geometric properties,
such as e.g. constant scalar curvature \cite{AppNewDev, EinsteinM}. This opens the possibility to interpret, in the context of GTD, models of DE as ordinary systems with constant thermodynamic interaction. In this work, we
find out a  new fundamental equations providing a constant thermodynamic curvature and analyze the corresponding cosmological model.
We describe these aspects in what follows.

\section{Dark energy fundamental equation from GTD}\label{cosmoGTD}

In this section, we derive our GTDF by starting from the consideration that the DE effects become relevant at a certain redshift during the history of the universe.
Such a remark is compatible with the hypothesis of a negative pressure proportional to the volume occupied by the GTDF itself.
In fact, if the volume occupied by the GTDF
 coincides with the Hubble sphere, it follows that effects due to DE can be neglected during matter and radiation dominated eras.
Given that, we may assume
\begin{equation}\label{EoSDE1}
P=-k\,V\,,
\end{equation}
where $k$ is a constant and $V$ is the volume of the universe. By keeping in mind that the pressure of a fluid is given by
\begin{equation}\label{PressDef}
P\equiv -\frac{\partial U(S,V)}{\partial V},
\end{equation}
where $U(S,V)$ is the internal energy of the system, using Eqs. (\ref{EoSDE1}) and (\ref{PressDef}), one can find the internal energy
\begin{equation}\label{uhjst}
U(S,V)=f(S)+k/2\,V^2\,,
\end{equation}
which represents a compatible fundamental equation, according to Eq. (\ref{EoSDE1}).
Despite of its apparent simplicity, the model presented in Eq. ($\ref{uhjst}$) leads to thermodynamic instabilities.
This is evident, since the internal energy is a sum of two functions: the first depending on the entropy only and the second one on the volume only.
Hence, constantly vanishing second order crossed derivatives occur, denoting thermodynamic instabilities.
Analogous conclusions can be inferred if $P=-f(V)$, where $f(V)$ is a generic function of $V$. In order to avoid unstable fluids, a different ansatz is proposed
here
\begin{equation}\label{EoSDE2}
P=-k\,V\,U(S,V)\,,
\end{equation}
which corresponds to a pressure which is proportional to both the volume and the internal energy. Integrating Eq. (\ref{EoSDE2}), one gets
\begin{equation}\label{fundEQ1}
U(S,V)=f(S)\,{\rm exp}\left(\frac{k}{2}\,V^2\right)\,.
\end{equation}
Plugging the fundamental equation (\ref{fundEQ1}) into the expression (\ref{gnat}) with $\Phi=U(S,V)$, $E^1=V$ and $E^2= S$ and calculating the scalar curvature,
we obtain that the constant thermodynamic curvature assumption is satisfied if and only if $f(S)$ is a polynomial of degree $2$ in $S$. Since we deal with an adiabatic
expansion (the universe is generally assumed to be expanding adiabatically \cite{hotncold}),  we are not interested in taking many orders for $S$, therefore we assume that
\begin{equation}\label{fund_eq}
{U}=U_0\exp\left(c_1	\,S+c_2\,{V}^2\right)\,,
\end{equation}
where $c_1$ and $c_2$ are constants to be fixed according to the physics of the problem.
Equation (\ref{fund_eq}) is the thermodynamic fundamental equation for our model, corresponding to the DE fluid according to our heuristic assumptions.
Using such equation, we can perform a GTD analysis of the system.
In the energy representation, the thermodynamic metric (\ref{gnat}) reads
\begin{equation}\label{gnatsysU}
 g^\natural_{U} = \frac{c_1^2}{2\,c_2\,V^2}\, {\rm d}S \otimes {\rm d}S+2\,\frac{c_1}{V}\,{\rm d}S \otimes {\rm d}V+\frac{\,(1+2\,c_2\,V^2)}{V^2}\,{\rm d}V \otimes {\rm d}V\,.
\end{equation}

We remark that, due to the invariance properties of the metric (\ref{gnat}), if we worked in the entropy representation (or in any other potential obtained from a total Legendre transformation
of $U$ or $S$) the results would have been the same. 
It is then just a matter of calculation to verify that the thermodynamic curvature in this case reads
\begin{equation}\label{RU}
R^\natural_U=-2\,.
\end{equation}
From the point of view of GTD, we interpret such a result noticing that  Eq. (\ref{fund_eq}) corresponds to a system with constant thermodynamic interaction.
Following the analysis of \cite{AppNewDev, AlejQuev, EinsteinM}, we wonder whether such systems can describe cosmological solutions.
We will show that our picture excellently fits the cosmological assumptions of the DE evolution.
Furthermore, the choice of constant curvature permits us to avoid singularities and phase transitions as $z\rightarrow0$.
The internal energy for our DE model as a function of the entropy and the volume is given by (\ref{fund_eq}).
From their thermodynamic definitions, it follows that
\begin{equation}\label{EoSDEbis}
T=c_1\,U_0\,{\rm e}^{{\it c_1}\,S+c_2\,V^2}\,,\qquad P=-2\,c_2\,V\,U_0\,{\rm e}^{{\it c_1}\,S+c_2\,V^2}\,,
\end{equation}
which, making use of (\ref{fund_eq}), can be expressed as
\begin{equation}\label{EoSDE2bis}
T=\,{\it c_1}\,U\,,\qquad P=-2\,c_2\,V\,U\,.
\end{equation}
From Eq. (\ref{EoSDE2bis}), we immediately notice that $c_1$ and $c_2$ must be positive constants, in order to have positive temperature and negative pressure. Alternatively, for the sake of clearness, introducing the DE energy density, i.e. $\rho_{DE}=U/V$, we can calculate the corresponding barotropic factor
\begin{equation}\label{omegaDE}
\omega_{DE}\equiv\frac{P}{\rho_{DE}}=-\,2\,c_2\,V^2\,.
\end{equation}
It is evident from Eq. ($\ref{omegaDE}$) that an increase on the volume is proportional (at constant energy $U$) to an increase of the pressure.
An immediate interpretation follows by considering that  at small volumes the negative pressure decreases, showing an expanding but non accelerating universe.
On the contrary, at larger volumes, the negative pressure contributes to the universe dynamics, eventually causing the observed late time acceleration.
The DE effects are therefore mimicked by our GTDF, by simply assuming the basic demands of GTD, with the recipe of constant thermodynamic interaction.

\section{Consequences in cosmology}
\label{FRWmodel}

Our GTDF can be used to develop a new cosmological model which can be adapted to a homogeneous and isotropic universe. To do so, we assume the spatially flat Friedmann-Robertson-Walker (FRW) metric
\begin{equation}\label{FRW}
ds^2=dt^2-a(t)^2\,(dr^2+r^2\,d\Omega^2)\,, \quad d\Omega^2\equiv d\theta^2 + {\rm sin}^2\theta\,d\phi^2\,,
\end{equation}
and a perfect fluid energy-momentum tensor of the form
\begin{equation}\label{Tab}
 T^{\alpha \beta}=(\rho_{t}+P)\, u^{\alpha} u^{\beta} - P g^{\alpha \beta}\,,
\end{equation}
where $\rho_t$ is the total energy density. It follows that the continuity equation reads
\begin{equation}\label{conserv1}
\dot\rho_t=-3\,H(t)(\rho_t + P)\,, \quad H(t)\equiv \frac{\dot a}{a}\,,
\end{equation}
where a dot denotes derivative with respect to time and $H(t)$ is known as the Hubble parameter. In our case $\rho_t$ is the sum of $\rho_{\rm matter}$ (the matter density) and $\rho_{DE}$, while $P$ is the DE pressure only, since matter is assumed to be pressureless. In Eq. (\ref{FRW}), $a(t)$ is the so-called {\it scale factor} of the universe (as a function of the cosmic time only),
which gives a measure of the linear scale of the universe, i.e. the entire volume should be proportional to $V=a(t)^3$.

In terms of the cosmological redshift, which is related to the coordinate time by
\begin{equation}\label{redshift}
\frac{dz}{dt} = -(1+z) H(z)\,,
\end{equation}
it follows that $a(t)=(1+z)^{-1}$, so that the volume of the universe can be written as \cite{iocalorespecifico}
\begin{equation}\label{Volume}
V(z) = \frac{1}{(1+z)^3}\,,
\end{equation}
for  adiabatic thermodynamic processes. With the definition of the redshift $z$ given in (\ref{redshift}), we can rewrite the energy conservation equation (\ref{conserv1}) as
\begin{equation}\label{conserv2}
\frac{d\rho_t}{dz}=\frac{3}{1+z}(\rho_t + P)\,.
\end{equation}
Recalling that $\rho_{\rm matter}=\Omega_m\,(1+z)^3$, and given the volume (\ref{Volume}),  from Eq. (\ref{fund_eq}) we find that the energy conservation equation (\ref{conserv2}) is directly satisfied. Moreover, the Einstein equations for the metric (\ref{FRW}) reduce to the well known Friedmann equations
\begin{eqnarray}\label{FReqs}
H^2&=&\rho_t\,,\label{FReqs1}\\
(1+z) \,H\,\frac{dH}{dz} - H^2&=&\frac{1}{2}\left(\rho_t+3P\right)\label{FReqs2}\,,
\end{eqnarray}
where we use units $8\,\pi\,G/3=1$ and we expressed the second equation directly in terms of the redshift $z$ using (\ref{redshift}). Assuming that the universe is undergoing an adiabatic expansion, which is a common assumption (in fact it can be considered as an isolated system, see \cite{hotncold}),
we can rewrite the fundamental Eq. (\ref{fund_eq}) in terms of the redshift and calculate the evolution of the thermodynamic DE quantities in terms of $z$.
Assuming a constant entropy, the DE density which corresponds to our GTDF reads
\begin{equation}\label{fund_eq2}
\rho_{DE}={\rm exp}\left(c_1 \,S_0+\, \frac{c_2}{(1+z)^6}\right)\,(1+z)^3\,,
\end{equation}
where we assumed that the volume is given by Eq. (\ref{Volume}). With Eq. (\ref{fund_eq2}) at hand, we can substitute into the first Friedmann equation
(\ref{FReqs1}) and define
\begin{equation}\label{Hubblez}
H(z)^2\equiv\rho_{\rm matter}+\rho_{DE}=\left[\Omega_m+{\rm exp}\left(c_1 \,S_0+\, \frac{c_2}{(1+z)^6}\right)\,\right](1+z)^3\,.
\end{equation}
We notice that from the flatness condition ($\Omega_m+\Omega_{DE}=1$ at $z=0$), it follows that ${\rm exp}\left(c_1 \,S_0+\, c_2\right)=1-\Omega_m$. Finally, with $H(z)$ given by (\ref{Hubblez}) and $\rho_t$ and $P$ defined according to the previous construction,
the second Friedmann equation (\ref{FReqs2}) is naturally satisfied. We can then calculate the barotropic factor (\ref{omegaDE}) as a function of $z$, which is easily rewritten as
\begin{equation}\label{omegaDEz}
\omega_{DE}=-\frac{2\,c_2}{(1+z)^6}\,,
\end{equation}
and suggests that we can recover the $\Lambda$CDM value $\omega=-1$ near $z=0$ by fixing the value of $c_2$ to be approximately $c_2\sim 0.5$.

\subsection{Energy conditions and adiabatic speed of sound}

To further investigate the robustness of our model, we can analyze the so called energy conditions (ECs),
which are relations that guarantee stability and causality of cosmological systems.
For a fluid as the one given in Eq. (\ref{Tab}), the ECs  are given by \cite{fregax}
\begin{eqnarray}\label{ECs}
\rho_t &\geq& 0, \quad \rho_t + P \geq 0 \quad {\rm (Weak\,\, Energy\,\, Condition)}\\
\rho_t + 3P &\geq& 0, \quad  \rho_t +P \geq 0 \quad {\rm (Strong\,\, Energy\,\, Condition)}\\
\rho_t - P &\geq& 0, \quad \rho_t +P \geq 0 \quad {\rm (Dominant\,\, Energy\,\, Condition)}
\end{eqnarray}

\begin{figure}\label{plotenergyconditions}
\begin{center}
\includegraphics[width=3.0in]{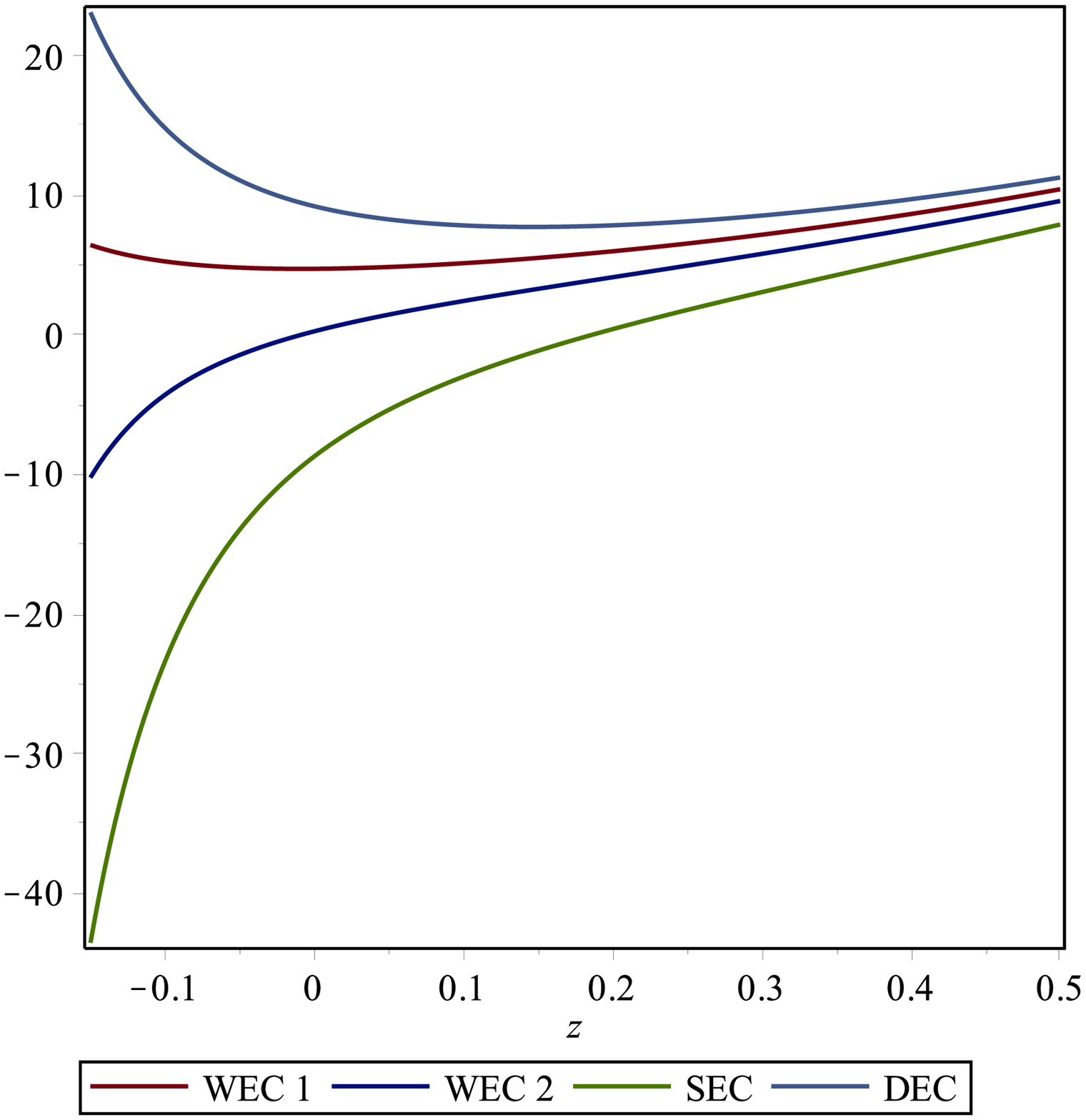}
\includegraphics[width=3.0in]{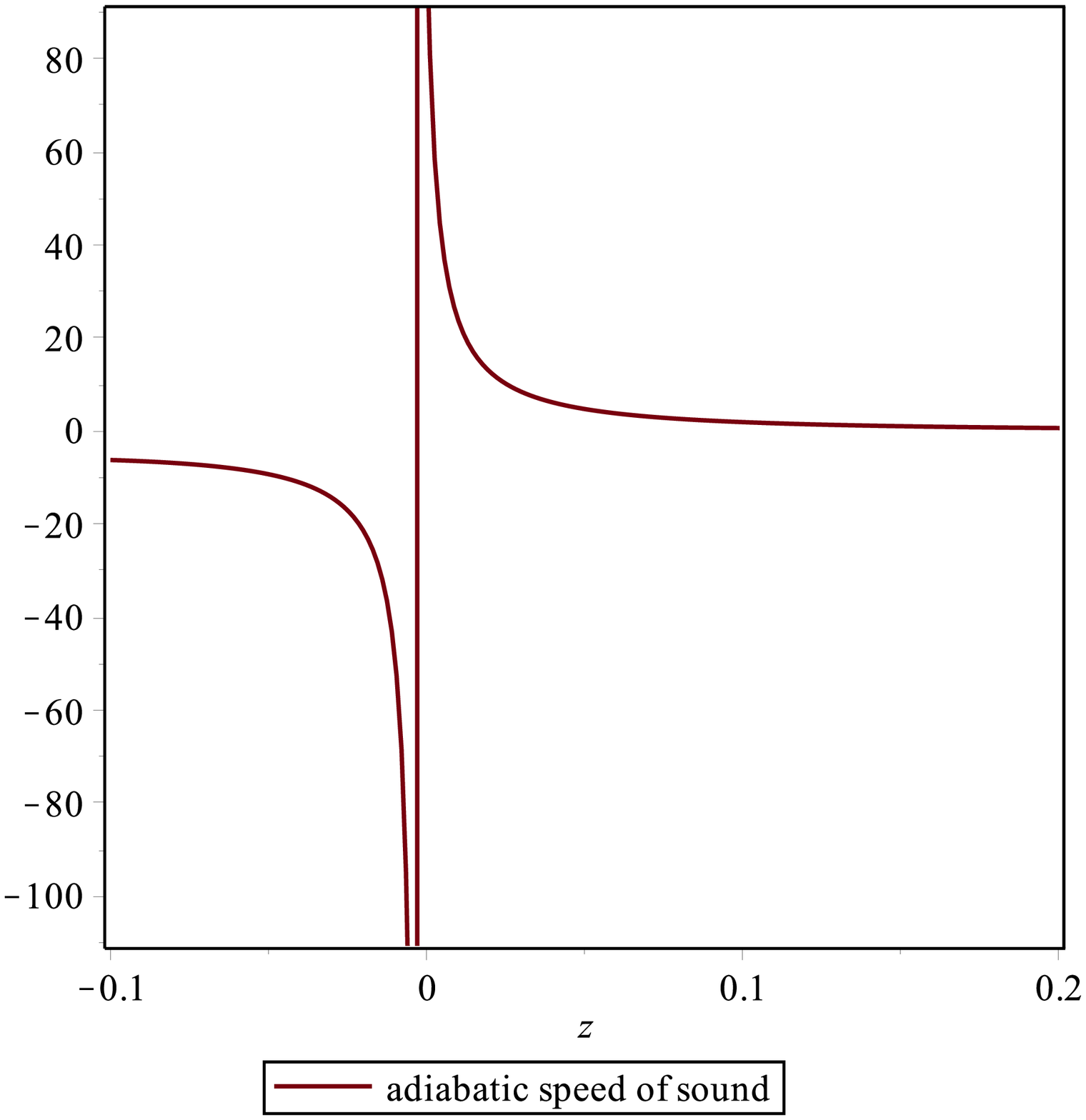}
\caption{The plot shows the evolution of the ECs over time (parametrized by the redshift $z$) for the standard value of $\Omega_m=0.25$ and for $c_2=0.5$ and $c_1=1/S_0$.
The red and blue lines indicate the WEC, the green line is the SEC and the cyan line
represents the DEC.}
\caption{The plot shows the evolution of the adiabatic speed of sound over the redshift $z$, with $\Omega_m=0.25$,  $c_2=0.5$ and $c_1=1/S_0$.}
\end{center}
\end{figure}

As we can see from Fig. 1, it turns out that all the ECs are satisfied in the past, i.e. for values of the reshift $z\geq0$,
while some of them (the second part of both the weak energy condition and strong energy condition) come to a point in the future evolution where they cease to be satisfied. However, this may be due to the assumption that $V=(1+z)^{-3}$, which exhibits a singularity at the future time, i.e. $z=-1$, being responsible of violating the ECs. In other words, our solution seems to approximate the behavior of the universe when $z> 0$, so that the singular and ill-defined future evolution is excluded from the region of interest.

On the other side, to guarantee the causality structure at early times, we investigate the adiabatic speed of sound,
related to pressure perturbations, i.e. $\delta P\approx c_S^2\delta\rho$, and given by \cite{oncemore}
\begin{equation}\label{speedofsound}
c_S^2\equiv\left(\frac{\partial P}{\partial \rho}\right)_S=\frac{6 \,{c_2} \left(2 {c_2}+(1+z)^6\right)\, e^{{c_1} {S_0}+\frac{{c_2}}{(1+z)^6}}}{(1+z)^{10} \left(\frac{3 \left[(1+z)^6-2 {c_2}\right] e^{{c_1}{S_0}+\frac{{c_2}}{(1+z)^6}}}{(1+z)^4}+3\, {\Omega_m} (1+z)^2\right)}\,,
\end{equation}
where in the last equality we have used Eq. (\ref{EoSDE2bis}) and the definition of the volume (\ref{Volume}).
To guarantee causality, it must be $c_S^2<1$, which is a natural result of our model, obtained for almost all $z>0$, although near $z=0$ a singularity occurs as due to the divergence in the adiabatic volume. A plot of $c_S^2$ with $c_2=0.5$ is given in Fig. 2.

\section{Reproducing the $\Lambda$CDM model as a low redshift limiting case}\label{LCDM}

We want to demonstrate that our model, based on assuming a GTDF fluid, reduces to $\Lambda$CDM when $z\ll1$. To understand this fact, let us expand $\rho_{DE}$ in a series around $z=0$, so that
\begin{equation}\label{rhodeespansa}
\rho_{DE}\approx (1 - \Omega_m)\left[1 +3 \,(1-2\,c_2)\,z+3\,(1+c_2+6\,c_2^2)\,z^2\right]+\mathcal{O}(z^3)\,.
\end{equation}
Using $P_{DE}|_{z=0}=-2\,c_2 \,(1 - \Omega_m)$, we obtain $\rho_{DE}\approx\rho_\Lambda\,\left[1+3\,(1+p_0)\,z+\dots\right]$,
where $\rho_\Lambda$ is the cosmological constant density. In other words, the cosmological model inferred from GTD is able to reproduce the universe dynamics, mimicking the role played by the cosmological constant through a geometrical fluid coupled, at a first approximation, to its own pressure \cite{th}. Moreover, the zeroth order mimics the cosmological constant, interpreting the $\Lambda$CDM model as a limiting case of a more general paradigm \cite{coodr,wei,ulti}.
This result turns out to be relevant to determine the meaning of the free parameters, which are related to the initial values of $\rho_m$ and $\rho_{DE}$. In particular, the coincidence problem, i.e. the awkward fact that both matter and DE magnitudes are comparable at late times \cite{babil}, is naturally alleviated by noticing that the GTDF reduces to $\rho_\Lambda$, as $z\rightarrow0$. These conditions simply define the values of constants in terms of the matter density, providing a description of the coincidence problem in terms of a setting of initial conditions.
In addition, since no vacuum energy cosmological constant is involved \emph{a priori},
we are able to alleviate the so called fine-tuning problem, dealing with the discrepancies between the theoretical and observational
values of the cosmological constant today \cite{oncemoreover}. Finally, looking at the Hubble rate Eq. ($\ref{Hubblez}$),
it is easy to notice that the DE term evolves in time by decreasing its contribution to the net energy budget as $z\rightarrow\infty$.
This is important because it permits to avoid discrepancies at higher redshift regimes, in which no modifications of the power spectrum is expected to occur. This agrees with the fact that at $z\gg 1$ observations indicate that DE does not significatively contribute to the power spectrum picks. In other words, at higher redshifts, our model is approximated by a pressureless matter fluid, as in the $\Lambda$CDM picture. The DE term then increases its contributions and dominates over matter at the so-called {\it transition redshift}, indicated as $z_{tr}$. In order to define $z_{tr}$ for our model, let us define the acceleration parameter, i.e.
\begin{equation}\label{q}
q=-1+\frac{(1+z)}{H}\frac{d H}{dz}=-1+\frac{3 \,\left((1+z)^6-2 {c_2}\right) e^{{c_1} {S_0}+\frac{{c_2}}{(1+z)^6}}+3 {\Omega_m} (1+z)^6}{2 (1+z)^6 \left(e^{{c_1} {S_0}+\frac{{c_2}}{(1+z)^6}}+{\Omega_m}\right)}\,,
\end{equation}
whose present day value  for our model reads
\begin{equation}\label{q0}
q_0=\frac{1}{2}-3\,c_2 (1-\Omega_m)\,,
\end{equation}
where we have used the flatness condition $e^{{c_1} {S_0}+{c_2}}=1-\Omega_m$. At a first order of approximation, one gets
\begin{equation}\label{zetsr}
z_{tr}\approx\frac{1}{9}\,\frac{3\,\Omega_m-2}{\Omega_m^2+\Omega_m-2}\,,
\end{equation}
which corresponds to $q=0$, i.e. the transition between the matter and DE dominated eras.
Furthermore, the rate of change of $q$ is measured by the {\it jerk parameter}, whose definition is
\begin{equation}\label{kj}
j=\frac{\ddot{ H}}{H^3}-3q-2=1+\frac{9\, {c_2} \left(2\, {c_2}+(1+z)^6\right) e^{{c_1} {S_0}+\frac{{c_2}}{(1+z)^6}}}{(1+z)^{12} \left(e^{{c_1} {S_0}+\frac{{c_2}}{(1+z)^6}}+{\Omega_m}\right)}\,.
\end{equation}
The present day value of $j$ for our model reads
\begin{equation}\label{jerko}
j_0=1+9 \,{c_2} (2 {c_2}+1) (1-\Omega_m)\,,
\end{equation}
where we used again the flatness condition. Comparing the estimated values for the acceleration and jerk parameters today and for $z_{tr}$ \cite{kom,wow,ancora}, we will be able to infer the cosmological bounds, which are confirmed in the observational limit, showing that our model is compatible with current data.

\section{Conclusion and perspectives}\label{Conclusion}

In this paper we investigated possible applications of thermodynamic geometry in order to infer new physically relevant informations in the field of cosmology. In particular, the core idea is that fluids with special geometric properties can also have peculiar thermodynamic behaviors. In particular in this work, we considered the paradigm of GTD and we derived a new fundamental thermodynamic equation corresponding to a fluid with constant thermodynamic curvature, since it has been speculated in previous works that systems with such geometrical property can show DE effects \cite{AppNewDev, AlejQuev, EinsteinM}. More generally, we proposed a technique to infer cosmological equation of states, able to describe the universe dynamics through geometric considerations. Afterwards, we demonstrated that it is possible to reproduce the effects due to DE by considering the GTDF and that fixing the free parameters of our model, the resulting fluid can effectively be responsible for the observed cosmic speed up. In fact, the predictions of our model well adapt to the cosmological dynamics both at present and past times, assuming as cosmological volume the adiabatic relation $V\propto {a}^{3}$. Furthermore, both the energy conditions and the adiabatic speed of sound are well behaved in the context of our model, in the region of interest for the generation of pressure perturbations. To better explain the main features of our model, we evaluated the acceleration and jerk parameters and the transition redshift in terms of the free parameters. Finally, we noticed that our paradigm reduces to $\Lambda$CDM as $z\rightarrow0$ and predicts a matter dominated era when $z\rightarrow\infty$. The fact that our approach extends the $\Lambda$CDM model gives in turn a possible explanation of the coincidence problem as a setting condition. Afterwards, the fine tuning problem is alleviated by assuming no cosmological constant fixed \emph{a priori} in the energy momentum tensor. Future efforts will be devoted to compare our results with observations using all the physical quantities evaluated here. Moreover, we plan to extend this work with additional forms of the volume $V$ and to further investigate the physical properties of fluids with special geometrothermodynamic properties.

\end{document}